# Chemically Meaningful Textualization Enables Explainable Validation of Metal-Organic Frameworks by Large Language Models

Guobin Zhao,[a] Xiao-Yan Li*[a]

a Department of Chemistry, National University of Singapore, 3 Science Drive 3, Singapore 117543, Singapore

* Corresponding author: xiaoyanli@nus.edu.sg

## Abstract

Computation-ready metal-organic framework (MOF) databases are essential for high-throughput screening, yet many reported crystal structures remain chemically unreasonable or disordered, compromising simulation fidelity. Existing validation approaches can identify non-computation-ready structures, but they often rely on heuristic rules, license requirement, or offer limited interpretability. Here, we show that large language models (LLMs) can serve as interpretable validators of MOF structures when crystallographic information is transformed into chemically meaningful text. By benchmarking nine descriptors, we find that successful LLM-based validation depends not on the amount of structural information alone, but on whether local coordination, framework connectivity, and chemical context are organized into a linguistically learnable representation. Fine-tuned LLMs using specialized descriptors (*mof2text*) achieve performance comparable to graph-based models in identifying unreasonable MOFs. Importantly, these models extend beyond black-box classification by generating diagnostic rationales for likely error sources, including abnormal bonding, connectivity, and charge states, as well as error-category predictions for annotated datasets. This work establishes chemically informed textualization as the key step that transforms LLMs from generic text models into practical and explainable tools for curating MOF databases.

## Introduction

Metal-organic frameworks (MOFs) are among the most intensively studied porous materials because of their structural diversity, tunable chemistry, and wide utility in gas adsorption and separation.[1-4] Their discovery has been increasingly accelerated by database-driven computational screening and machine learning (ML), where crystallographic information files (CIFs) serve as the foundation for simulations, screening, and property prediction. However, many experimentally reported MOF structures contain disorder, missing atoms, charge imbalance, or chemically unreasonable coordination environments.[5-7] Such structural defects can severely distort computed properties and compromise the reliability of high-throughput workflows.[8, 9]

Several validation strategies have therefore been developed, including geometry-based heuristics, oxidation-state analysis, and graph-based ML models. Existing efforts in MOF database curation, such as the Computation-Ready Experimental MOF Database (CoRE MOF DB)[10, 11] and the MOSAEC MOF database,[12] can already identify many unreasonable structures. However, important limitations remain: rule-based approaches often lack high precision or restricted access,[13, 14] whereas graph-based classifiers usually provide limited interpretability.[11, 15] In practice, identifying that a structure is unreasonable is only the first step; effective refinement further requires an interpretable diagnosis of the underlying defects.

Large language models (LLMs) offer a potentially attractive route to bridge prediction and explanation.[16, 17] In reticular chemistry and materials science, LLMs have been explored for literature mining, information extraction, and human–AI-assisted design.[18-24] However, their application to crystallographic validation remains largely unexplored. A central bottleneck is that crystal structures are not naturally expressed in a form that general-purpose LLMs can readily interpret.[25-28] For MOFs, structural validity depends simultaneously on local coordination chemistry, long-range framework connectivity, and chemically consistent charge distribution. Simply converting CIFs into raw text is therefore unlikely to provide an adequate representation.[29]

In this study, we investigate whether LLMs can be adapted for MOF validation through chemically informed textualization of crystal structures. We first show that unreasonable MOF structures can induce substantial deviations in computed energetic, electronic properties, and adsorption properties, highlighting the practical need for reliable validation. We then benchmark nine descriptors spanning non-text, semi-text, and full-text formats to identify which structural representations allow LLMs to internalize crystallographic validity. Finally, we demonstrate that, once paired with chemically meaningful descriptors, LLMs can move beyond binary classification to provide interpretable rationales and error-type predictions for structurally unreasonable MOFs. This work clarifies how MOF structures should be textualized for explainable validation and diagnosis by LLMs, thereby establishing a practical route toward AI-assisted curation of computation-ready MOF databases.

## Results and Discussion

We first examined whether structurally unreasonable MOFs distort computed properties relevant to screening and simulation. Structural disorder and missing ions cause clear shifts in the electronic energy and band gap predicted by the graph neural network (GNN) (**Figure 1a, b)**, showing that even modest crystallographic inconsistencies would substantially compromise the reliability of ML-based predictions. They also strongly affect the gas adsorption simulations (**Figure 1c, d)**. For structures lacking charge-balancing ions, $CO_2$ uptake is markedly overestimated due to an enlarged accessible pore volume, as supported by the void fraction (VF) analysis (**Figure S2**), and by altered charge distribution, which in turn leads to major deviations in predicted gas adsorption and separation. Energy and band gap data were predicted by PACMAN,[30] whereas the adsorption data were obtained from Grand Canonical Monte Carlo (GCMC) simulations for mixture adsorption,[31] with computational details provided in **Supporting Information Section S1 and Figure S1**. Taken together, these results suggest that structural validation is essential for constructing physically meaningful, computation-ready MOF datasets.

Experimental $H_2$ adsorption isotherms[32, 33] are reproduced much more faithfully by structurally corrected MOFs than by the corresponding raw MOFs, as shown in **Figure 1e, f**. Specifically, the unreasonable raw structures with missing H atoms ("h", VEWKUF) and structure disorder ("disorder", WAFKAQ) also exhibit underestimation and overestimation of the gas uptake, respectively. An additional case, PCN-233, which simultaneously contains both "h" and "disorder" defects, also shows overestimated uptake (**Figure S3**). These results further emphasize that effective structural validation is essential not only for constructing physically meaningful, computation-ready MOF datasets but also for bridging the gap between theoretical simulations and experiments.

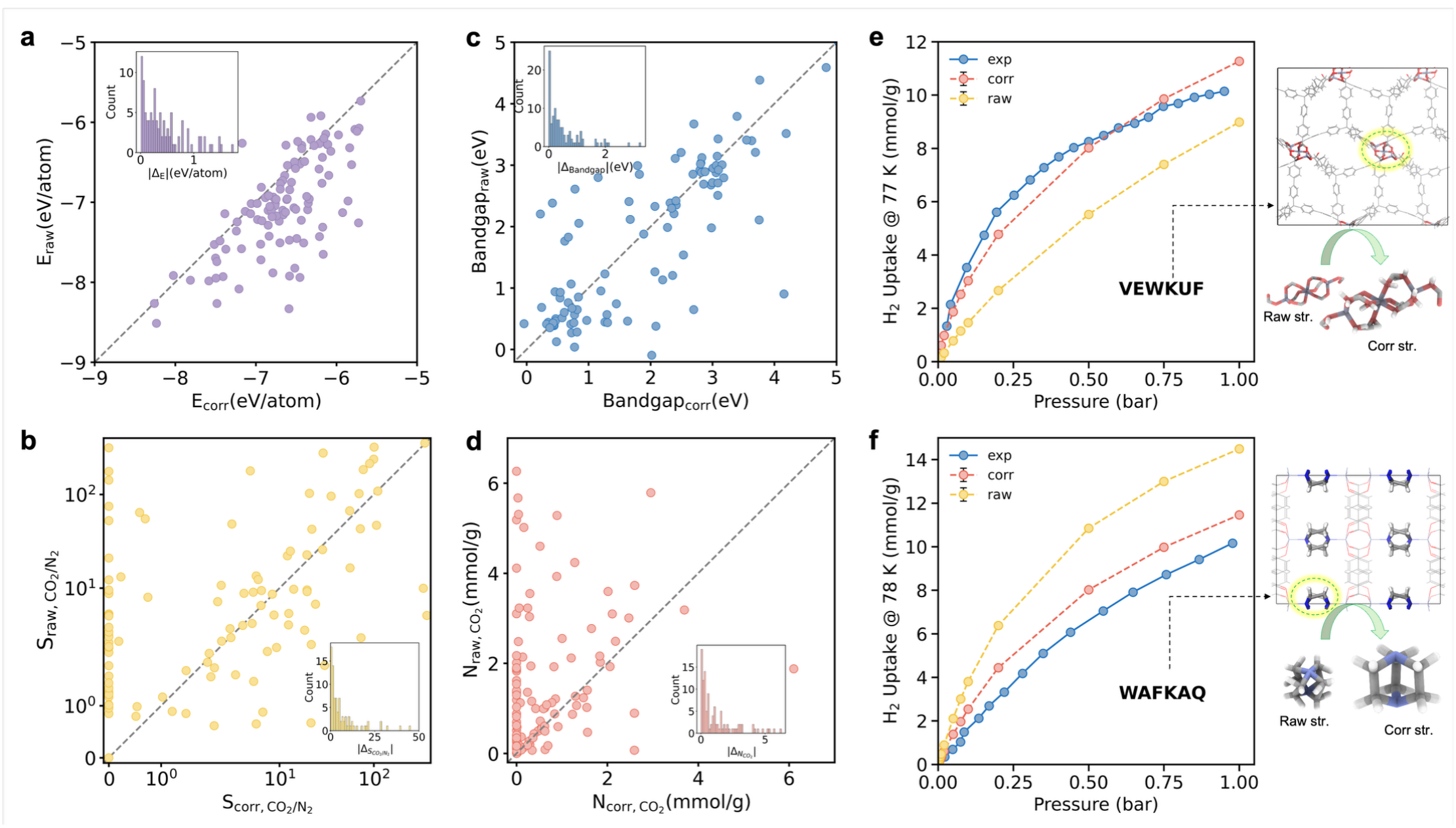

**Figure 1.** Comparison between raw and corrected (corr) MOF structures in terms of (a) electronic energy, (b) band gap, (c) $CO_2/N_2$ selectivity, and (d) $CO_2$ uptake, where the insets show the distribution of absolute differences. Comparison of experimental and simulated $H_2$ adsorption isotherms for raw and corrected structures for (e) VEWKUF and (f) WAFKAQ.

We then investigated whether LLMs can be used for MOF structural validation and diagnosis. Most descriptors (crystal_text_LLM,[34] robocry[35]) currently used in LLM-based materials studies were originally developed for inorganic crystals and therefore do not fully capture the structural informatics of complex MOFs, where validity depends not only on local coordination environments, but also on framework connectivity, topology, chemically meaningful building units, and charge balance (the detail can be found in **Supporting Information Section S2.2**). We therefore benchmarked nine descriptors spanning non-text, semi-text, and full-text, and compared their performance in both pre-trained (PT) and fine-tuned (FT) LLMs using four evaluation metrics (**Figure 2**; **Supporting Information Section S3**). We define descriptors consisting solely of chemical element and number as non-text descriptors. Descriptors that additionally contain string-based components are defined as semi-text descriptors, whereas descriptors expressed in natural language are defined as full-text descriptors. The corresponding representations for a representative crystal structure (e.g., IRMOF-1) are detailed in **Supporting Information Section S2.2**.

For PT-LLMs, *local_env* descriptor, which encodes local chemical environments in MOFs, generally performs best, whereas descriptors containing crystallographic inputs or other more explicit geometric detail offer little advantage. Notably, even simple compositional inputs perform comparably to, or better than, more detailed structural encodings (atomic coordination, space group and so on), consistent with prior observation in synthesizability prediction.[25] This suggests that general-purpose LLMs do not readily extract chemically useful information from raw crystallographic text alone. More broadly, these results reveal that native LLM knowledge is insufficient for MOF structural validation and that representation design is already a limiting factor.

After fine-tuning, descriptors based on full-text representations,[36] particularly *robocry*[35] and *mof2text* (**Figure S10**), consistently show superior predictive performance, whereas non-text or semi-text representations remain markedly less effective. This trend indicates that successful LLM-based validation depends not simply on the amount of structural information provided, but on whether that information is reorganized into a chemically meaningful and linguistically learnable form. The improved performance of *local_env* relative to *struc2str* further suggests that LLMs benefit when crystallographic information is reformulated into a more chemically legible representation, rather than passed through as a direct geometric transcription. Similarly, *cif_p1* and other inefficiently compressed inputs (*crystal_text_llm* and *slices*) exhibit excessively long token lengths (**Figure S9**), increasing computational cost without corresponding gains in predictive accuracy. Notably, *mof2text* encodes MOF-relevant features, such as topology, charge-related information, and framework building units, while avoiding redundant atom-by-atom description of the periodic structure. Together, these findings identify that chemically meaningful textualization, rather than raw structural completeness, is the enabling step for LLM-based MOF validation.

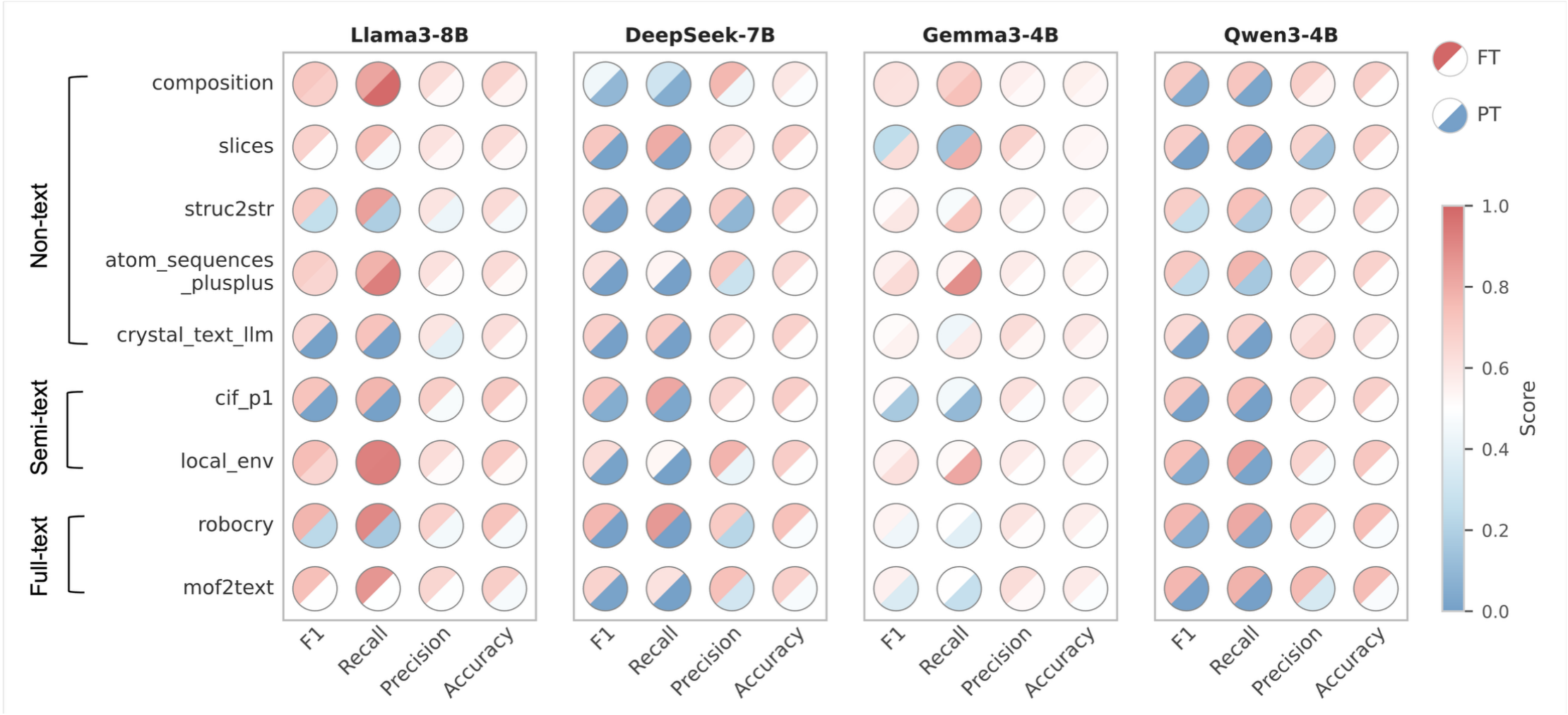

**Figure 2.** Benchmark of four representative LLMs (Llama3-8B, DeepSeek-7B, Gemma3-4B, and Qwen3-4B) across nine structural descriptors, comparing pre-trained (PT, lower-left half circle) and fine-tuned (FT, upper-left half circle) models using four metrics.

We further compared the fine-tuned LLM-based framework with established validation methods and existing ML models (**Table 1**; **Supporting Information Section S4**). Geometry-based approaches, such as Chen–Manz (bond-order-based method)[9] and MOFChecker (geometry-based method),[13] show limited overall accuracy. MOSAEC (metal oxidation-based method), which is based on metal oxidation-state analysis, remains highly effective but depends on commercial access to the Cambridge Structural Database (CSD). PMTransformer, which integrates both local and global features, emerges as the top-performing model overall. Text-based descriptors excel at integrating local (unique unit, bond length and atomic charges) and global features (chemical formula, space group, dimension, topology and lattice parameters), whereas Transformer-based models (LLMs) can effectively capture long-range dependencies. Importantly, fine-tuned LLMs paired with chemically meaningful descriptors achieve performance comparable to that of graph-based models (MOFClassifier, CGCNN) and clearly outperform geometry-based (MOFChecker) methods. Among them, the FT-Llama model (*robocry*) yielded the highest recall, reflecting a more stringent threshold for labeling structurally reasonable MOFs. The FT-Qwen3 (*mof2text*) achieved the highest accuracy, effectively balancing precision and recall. These comparisons demonstrate that once MOF structures are represented in a chemically meaningful form, fine-tuned LLMs become competitive and interpretable validators of MOF databases.

**Table 1**. Performance of various methods and models for the structural validation of MOFs. Evaluated metrics include accuracy (Acc), precision (Prec), true positive rate or recall (TPR), and F1 score.

| **Method** | **Accuracy** | **Precision** | **Recall** | **F1** |
|---|---|---|---|---|
| Chen-Manz[a] | 0.587 | 0.699 | 0.290 | 0.410 |
| MOFChecker[a] | 0.623 | 0.812 | 0.310 | 0.449 |
| MOSAEC[b] | 0.902 | 0.946 | 0.851 | 0.896 |
| MOFClassifier[a] | 0.764 | 0.755 | 0.775 | 0.765 |
| SETC[b] | 0.869 | 0.926 | 0.800 | 0.858 |
| CGCNN[c] | 0.807 | 0.771 | 0.884 | 0.824 |
| PMTransformer[c] | 0.866 | 0.863 | 0.870 | 0.867 |
| LLM-prop[c,e] | 0.628 | 0.597 | 0.893 | 0.716 |
| Llama-3-8b[c,d] | 0.727 | 0.679 | 0.900 | 0.774 |
| Qwen-3-4b[c,e] | 0.757 | 0.762 | 0.782 | 0.772 |

a The method (model) used in this work is derived from the literature.[9, 11, 13] b The data from literature.[14, 15] c Models trained or fine-tuned in this work and their results on the test set. d *robocry*. e *mof2text*.

Beyond identifying unreasonable MOFs, we next asked whether LLMs can also diagnose their likely structural origins. The t-distributed stochastic neighbor embedding (t-SNE) shows that, for given raw (analogous, simplified) crystallographic data formats, such as *cif_p1*, *crystal_text_llm*, and *atom_sequences_plusplus*, the embeddings of reasonable and unreasonable MOFs largely overlap (**Figure 3a**). By contrast, *mof2text*, *robocry*, and *local_env* produce much clearer separation. Similar distributions were observed for other tested LLMs (**Figures S19-S21**). These results indicate that descriptors that preserve chemically interpretable framework information not only improve validation accuracy but also organize the embedding space in a way that more accurately reflects structural reasonableness. Consequently, *mof2text* not only demonstrates superior classification performance but also makes the generation of interpretable rationales derived from t-SNE discrepancies possible.

We therefore used PT-LLMs to generate the rationales for the structurally unreasonable MOFs. Using *mof2text* as input, Qwen3 most frequently identified defects related to "bonding" and "connectivity" (**Figure 3b**; **Table S2**), whereas charge-related anomalies were captured more often when PACMAN-derived atomic charge information was included in the textual description. Rather than producing generic explanations, the models repeatedly mapped unreasonable structures onto chemically meaningful categories. This behavior suggests that once structural information is textualized in an appropriate form, LLMs can associate classification outputs with chemically plausible defect patterns. A case with multi errors as shown in Figure S16.

The t-SNE plot of unreasonable MOFs colored by LLM-generated rationales further supports this interpretation (**Figure 3c**). In particular, MOFs assigned charge-related rationales occupy a more concentrated distribution in the embedding space, while the overall distribution shows discernible local clustering rather than random scatter. We then compared the distribution of LLM-generated rationales with four manually labeled defect classes from the literature, namely "h" (missing hydrogen atoms), "disorder" (crystallographic disorder), "charge" (charge imbalance), and "other" (other error categories) (**Figure 3d**). Meaningful overlap is observed, although strict one-to-one correspondence is neither expected nor required, because a single crystallographic defect often triggers multiple coupled chemical inconsistencies. For example, missing H atoms can simultaneously result in unreasonable coordination environments, unsaturated bonding, and anomalous atomic charges. Similarly, structural disorder can perturb bonding, connectivity, and charge distribution simultaneously. Consistent with this view, comparison with rule-based diagnosis (**Figures S17 and S18**) shows that defects labeled as "overcoordinated_n" are dominantly associated by the LLMs with connectivity, symmetry, coordination, and bonding, whereas "high_charges" is linked to charge, bonding, and connectivity. Therefore, errors from coarse human annotation or rule-based labeling typically contribute to, or propagate into, other error types, From LLM generation, capture not only isolated defect labels but also the broader chemical consequences of unreasonable structures.

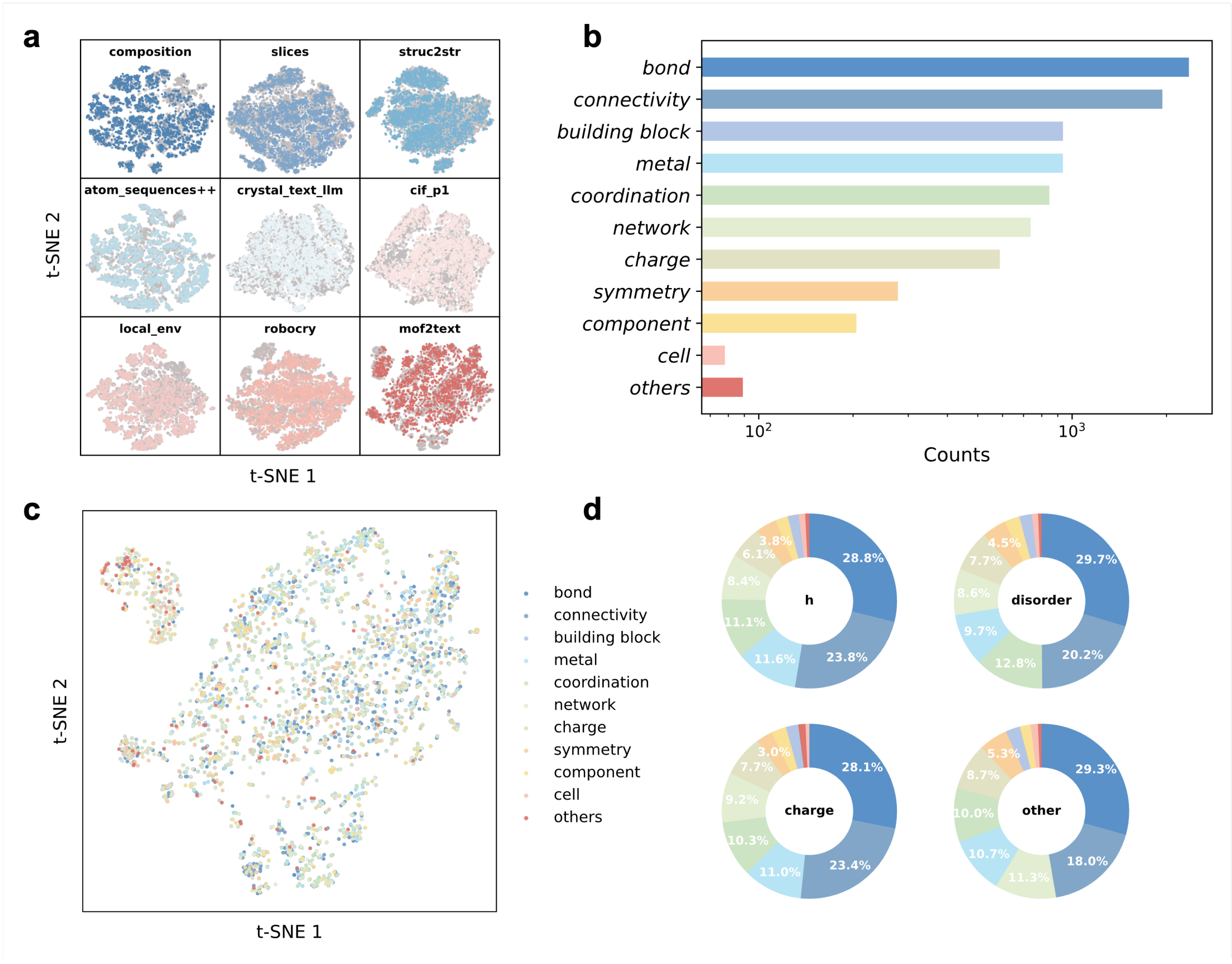

**Figure 3.** (a) t-SNE comparison of reasonable MOFs represented by different descriptors and unreasonable MOFs (black-edged white points). (b) Distribution of LLM-identified reasons for structurally unreasonable MOFs. (c) t-SNE map of unreasonable MOFs colored by LLM-generated reasons. (d) Pie chart of the correspondence between LLM-generated errors and human-annotated error types. Results in (b)-(d) were generated using *mof2text* as input with the Qwen3-4B model.

To further investigate the capability of fine-tuned LLMs to reason about the underlying causes of unreasonable MOFs, we employed a manually labeled four-class dataset (**Figure 3d**). As shown in **Figure 4a**, given unreasonable MOFs with error types, we evaluated the performance of FT-LLM on a manually curated dataset (FT-LLMs-R).

On the test set, the FT-Llama3-R model can correctly identify the error types of approximately 70% of unreasonable MOFs, while only about 12% of the predictions are incorrect (**Figure 4b–4d**). For "h" and "charge", more than half of the misclassifications arise from either predicting an alternative error type or missing one correct label in multi-error structures. These results suggest that, when supplied with domain-specific representations and curated annotations, fine-tuned LLMs can move beyond binary validation toward chemically informed error categorization.

At the same time, the four-class classification framework remains an intentionally simplified approximation of the full defect space. It cannot precisely localize the erroneous atoms or fully specify the root cause of each defect, both of which will be important for a future automated correction framework. For instance, the absence of an organic linker framework represents a non-MOF-like structural failure mode that can be recognized by PT-LLMs, yet is not exhaustively described by the current four-class annotation scheme. More broadly, classification alone is insufficient to recover the complete hierarchy of defect origin, propagation, and structural consequence. Future integration of broader MOF literature, curated defect corpora, and domain knowledge may therefore enable a more general workflow that combines validation, rationale generation, and autonomous refinement of unreasonable MOF structures.[37, 38]

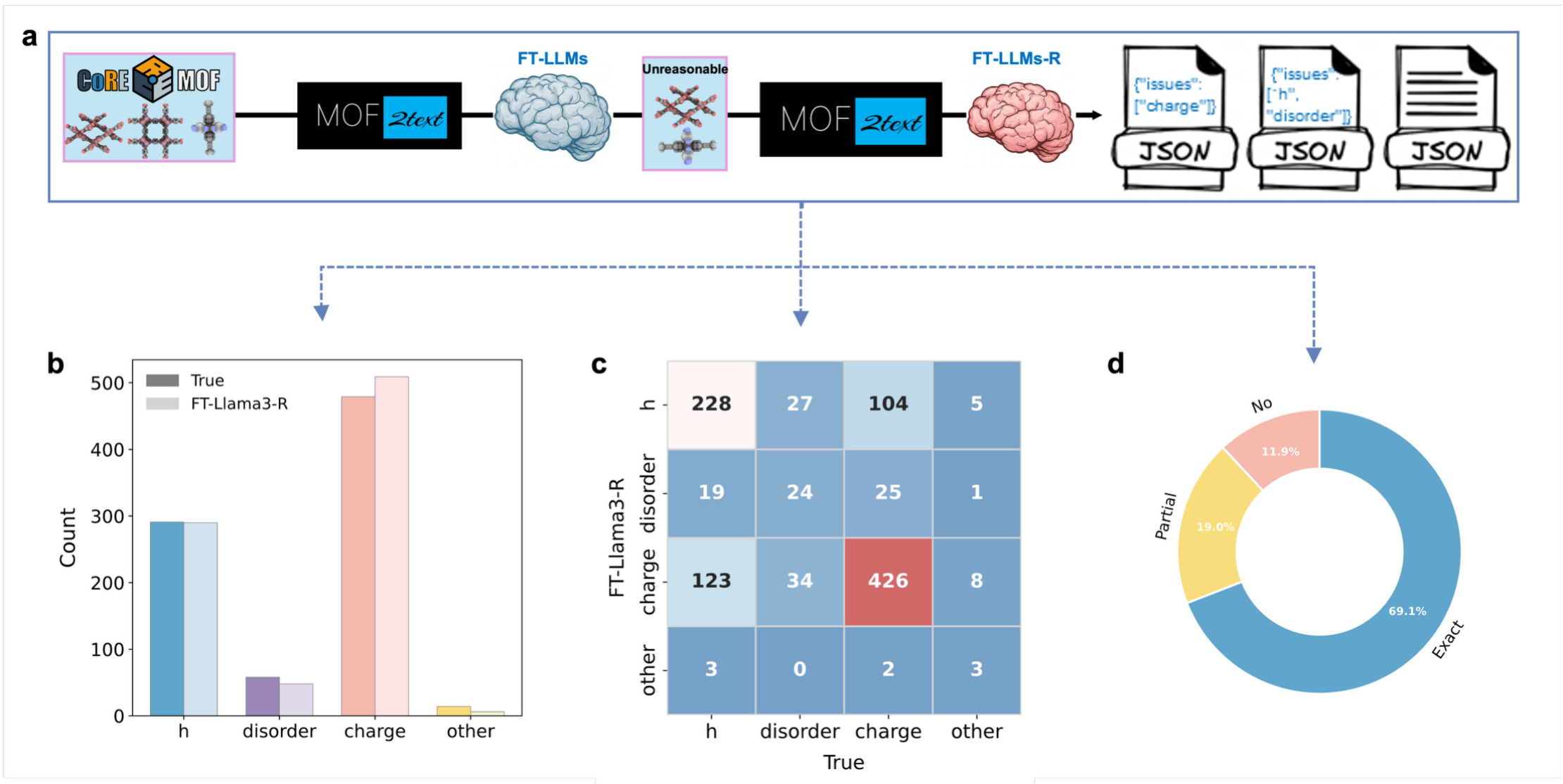

**Figure 4.** (a) Workflow of fine-tuned-LLMs for error reason prediction of MOFs. (b) Distribution of error types in unreasonable MOFs determined by manual inspection (True) and predicted by fine-tuned Llama3-8B (FT-Llama3-R). (c) Confusion matrix for the four-class classification of MOF error types, comparing the predictions of the fine-tuned Llama3-8B (FT-Llama3-R) against manual inspection (True). (d) Performance of FT-Llama3-R predictions. Since a single MOF may contain multiple errors, the results are categorized as: Exact (all errors matched), Partial (a part of errors

matched), and No match (no error types matched). All data in this figure are from the test set. The data for other models are shown in **Figures S22-24**.

## Conclusion

In conclusion, we show that LLMs can be adapted for explainable validation of MOF structures when crystallographic information is transformed into chemically meaningful text. Given the structural complexity of MOFs, reliable validation is necessary before theoretical simulation and high-throughput screening. Across nine descriptors, we demonstrate that LLM performance is governed not simply by the quantity of structural information, but by whether local coordination, framework connectivity, and chemical context are organized in a linguistically learnable form. Specialized descriptors such as *mof2text* and *robocry*, therefore, function as enabling representations rather than mere alternative encodings.

Fine-tuned LLMs achieve validation performance comparable to graph-based models and extend beyond black-box classification by providing chemically meaningful rationales and error-type predictions. These outputs provide actionable guidance to identify likely origins of unreasonable structures and may facilitate subsequent manual or automated refinement. More broadly, this work establishes that the value of LLMs in MOF science lies not only in prediction but also in their ability to couple prediction with interpretable diagnosis. Chemically informed textualization thus provides a practical route to transforming general-purpose language models into explainable assistants for curating computation-ready MOF databases.

## Acknowledgements

X.-Y. L. gratefully acknowledges the support from the National University of Singapore Start-Up Grant (grant A-0010269-00-00). We acknowledge the high-performance computing resources supported by NUS Information Technology (NUS-IT) in Singapore.

## Conflicts of Interest

The authors declare no conflicts of interest.

## Data Availability Statement

Detailed information is available in the Supporting Information of this article. The mof2text tool, scripts, machine learning models, and fine-tuned LLMs used in this work are available on GitHub at https://github.com/sxm13/MOF2Text and https://github.com/sxm13/MOFClassifier2. The dataset, embeddings, descriptors, fine-tuned-LLMs, and other raw data are available on Zenodo at https://zenodo.org/records/19144569. Access to the pre-trained and fine-tuned LLMs is publicly available for non-commercial use through Hugging Face at https://huggingface.co/ and https://huggingface.co/Sxm13/MOFClassifier2.

## Table of Contents

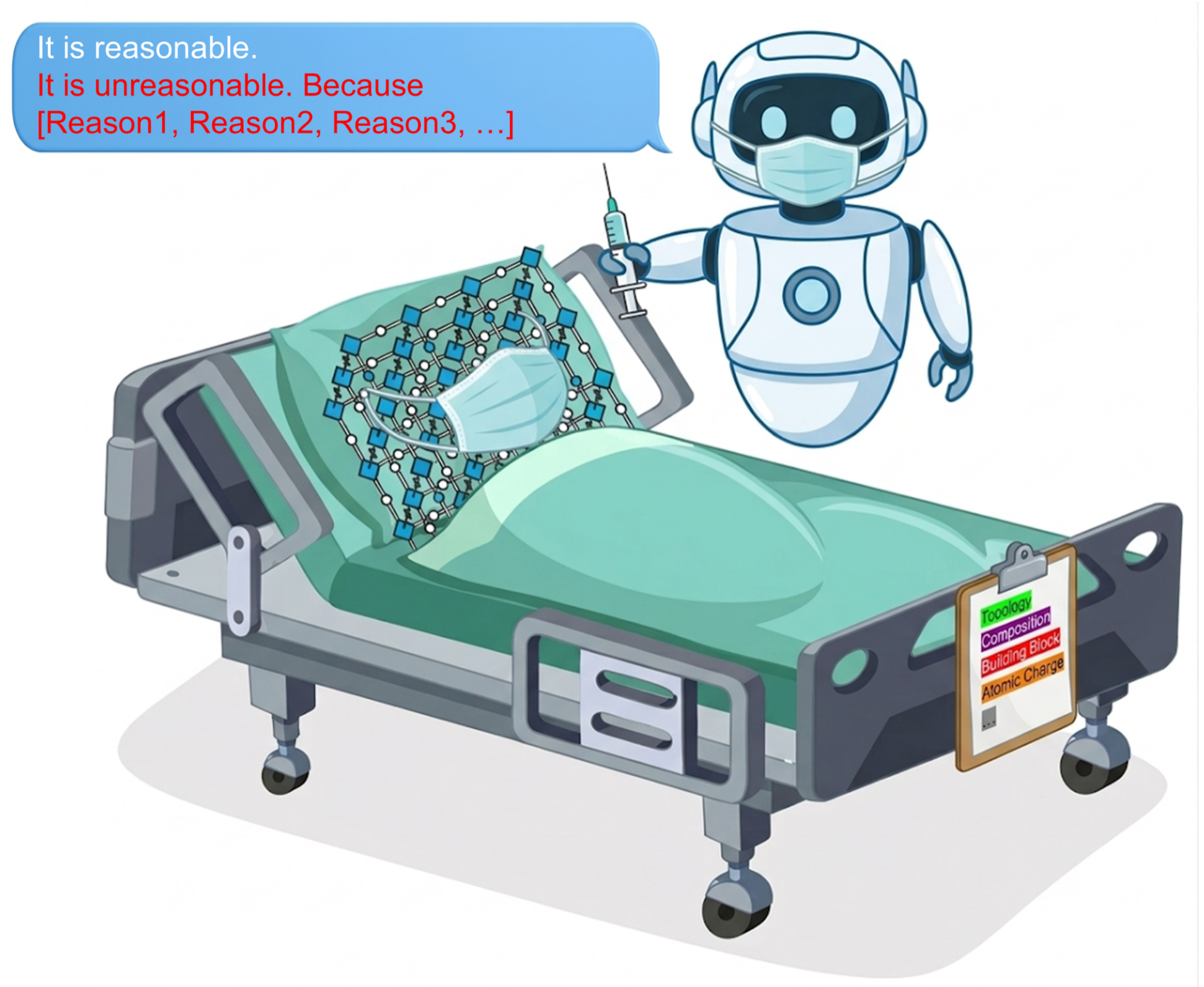